\documentclass[prd,
superscriptaddress,showpacs,nofootinbib,%
tightenlines ]{revtex4}
\usepackage{epsfig}
\usepackage{amssymb}
\usepackage{bbold}

\newcommand{\ben}{\begin{displaymath}}
\newcommand{\een}{\end{displaymath}}
\newcommand{\be}{\begin{equation}}
\newcommand{\ee}{\end{equation}}
\newcommand{\bea}{\begin{eqnarray}}
\newcommand{\eea}{\end{eqnarray}}

\usepackage{color}

\begin{document}
\title{Wilsonian renormalization group and the Lippmann-Schwinger equation with a multitude of cutoff parameters}
\author{E.~Epelbaum}
 \affiliation{Institut f\"ur Theoretische Physik II, Ruhr-Universit\"at Bochum,  D-44780 Bochum,
 Germany}
\author{J.~Gegelia}
\affiliation{Institute for Advanced Simulation, Institut f\"ur Kernphysik
   and J\"ulich Center for Hadron Physics, Forschungszentrum J\"ulich, D-52425 J\"ulich,
Germany}
\affiliation{Tbilisi State  University,  0186 Tbilisi,
 Georgia}
\author{U.-G.~Mei\ss ner}
 \affiliation{Helmholtz Institut f\"ur Strahlen- und Kernphysik and Bethe
   Center for Theoretical Physics, Universit\"at Bonn, D-53115 Bonn, Germany}
 \affiliation{Institute for Advanced Simulation, Institut f\"ur Kernphysik
   and J\"ulich Center for Hadron Physics, Forschungszentrum J\"ulich, D-52425 J\"ulich,
Germany}

\begin{abstract}
The Wilsonian renormalization group approach to the Lippmann-Schwinger
equation with a multitude of cutoff parameters is introduced.  
A system of integro-differential equations for the cutoff-dependent
potential is obtained. As an illustration, a perturbative solution of
these equations with two cutoff parameters for  
a simple case of an $S$-wave low-energy 
potential in the form of a Taylor series in momenta is
obtained. The relevance of the obtained results for the effective field
theory approach to nucleon-nucleon scattering is discussed.
 
\end{abstract}

\maketitle

\section{Introduction}
\label{introduction}

The chiral effective field theory (EFT) approach
to few-nucleon systems~\cite{Weinberg:1990rz,Weinberg:1991um} has
attracted much attention during the past two and a half decades. 
The problem of renormalization and power counting in this framework
turned out to be highly nontrivial and caused controversial
debates in the community. A number of formulations alternative to
Weinberg's original proposal have been suggested to resolve the issue
of renormalization, see 
Refs.~\cite{Bedaque:2002mn,Epelbaum:2008ga,Birse:2009my,Epelbaum:2012vx,Machleidt:2011zz,Valderrama:2016koj} 
for review articles. In our recent paper \cite{Epelbaum:2017byx} we
have compared a subtractive renormalization  approach with the Wilsonian
renormalization group (RG) approach \cite{Birse:1998dk,Harada:2010ba}  in the context of  the EFT for the
two-nucleon system close to the unitary limit. In particular, within the subtractive scheme, we
have identified the choices of renormalization conditions corresponding to
the Kaplan-Savage-Wise (KSW) \cite{Kaplan:1998tg}, see also Refs.~\cite{vanKolck:1997ut,vanKolck:1998bw}, and Weinberg
\cite{Weinberg:1990rz} power counting schemes. The
standard Wilsonian RG method with a single cutoff scale is, on the other hand, only
compatible with the KSW counting scheme. We argued that this
mismatch is caused by the too restrictive formulation of the Wilsonian
RG approach in its conventional form, which does not take into account  the
full freedom in the choice of renormalization conditions in EFT. This
is the origin of the often made (incorrect, see
Ref.\cite{Epelbaum:2017byx}) statement that the Weinberg
power counting scheme for two-nucleon scattering corresponds to the expansion around a  trivial
fixed point.

In the Wilsonian RG approach one integrates out
degrees of freedom with energies higher than some cutoff scale and
systematically 
exploits the cutoff-parameter dependence of coupling constants to
ensure that physics at energies below the cutoff scale remains
unchanged \cite{Wilson:1973jj}. In contrast, the Gell-Mann-Low RG
equations determine the dependence
of various quantities on the \emph{scale(s) of renormalization} \cite{GellMann:1954fq}.   
In renormalizable (in the traditional sense) theories only logarithmic divergences contribute
to the renormalization of the coupling constants and, therefore, there is a direct
correspondence between the two approaches.  On the other hand, in EFTs with
non-renormalizable interactions, power-law divergences have to be
taken care of and the direct link between the two 
RG equations is lost. Notice further that in theories with more than
one coupling constant, as it is the case in EFTs, each coupling
is attributed its own renormalization scale. In the Wilsonian approach
one usually introduces a single cutoff scale and studies how
various parameters of a theory depend on it. However, in certain cases
such as e.g.~the few-nucleon problem in chiral EFT, it is advantageous
to exploit the freedom of choosing several  
renormalization scales independently \cite{Gegelia:1999ja,Epelbaum:2017byx}. 

In this paper we fill this gap and generalize the Wilsonian RG analysis of
low-energy two-particle scattering in the framework of 
the Lippmann-Schwinger (LS) equation, pioneered in
Ref.~\cite{Birse:1998dk}, by introducing a multitude of cutoff
parameters. We obtain a system of integro-differential RG equations
describing the dependence of the potential on several cutoff
scales. As an application, we study a perturbative solution of the
obtained system of equations for the case of two cutoff parameters by
making an ansatz for the potential in the form of a Taylor series
expansion in powers of momenta. We demonstrate that the resulting
potential indeed obeys the Weinberg power counting for the 
choice of renormalization conditions suggested in
Ref.~\cite{Epelbaum:2017byx}. 

Our paper is organized as follows. In the next section we derive  the system
of RG equations for the case of several cutoff parameters.  In section~\ref{RG2C}, 
we present the perturbative solution of this system of
equations and discuss the obtained results in the context of EFT for
two-nucleon scattering. Finally, our main findings are briefly
summarized in section~\ref{sec:summ}.

\section{The  Lippmann-Schwinger equation with  a multitude of cutoff parameters}
\label{RGEqs}

To introduce a multitude of cutoff parameters and derive the
corresponding system of RG equations we start with the fully off-shell LS equation
\begin{equation}
T({\bf p},{\bf q},k)=V({\bf p},{\bf q},k)+\int \frac{d^3{\bf  l}}{(2\pi)^3} \, V({\bf p},{\bf l},k) G(k, {\bf l})\,T({\bf l},{\bf q},k),
\label{eq1}
\end{equation}
where $G(k,{\bf l})= 2 m/(k^2-{\bf l}^2+i\,\epsilon)$ is the
nonrelativistic two-particle Green's function, and $k^2/m$ is the
kinetic energy in the centre-of-mass frame. 
We assume that the low-energy dynamics of the system at hand is 
describable in the framework of the non-relativistic Schr\"odinger theory, 
i.e. that the underlying potential $V({\bf p},{\bf q},k)$ is non-singular and well-behaved 
in the quantum mechanical sense.  
We regard  Eq.~(\ref{eq1}) as an ``underlying'' model and follow the
philosophy of Wilson's renormalization group approach.
Specifically, we aim at
integrating out the high-momentum modes by introducing 2N cutoffs 
$\Lambda_1, \bar\Lambda_1,\Lambda_2, \bar\Lambda_2,\ldots, \Lambda_N,
\bar\Lambda_N $ such that the off-shell  
amplitude remains unchanged at low-energies. 
While in all practical applications one considers Hermitean cutoff potentials, corresponding to $\Lambda_i = \bar\Lambda_i$, to keep our resulting equations in the 
most general form we do not impose this condition in our derivation.
We start by writing the potential $V({\bf p},{\bf q},k)$ as a sum of
various contributions (the choice of which depends on the particular
problem one is dealing with) 
\begin{eqnarray}
V({\bf p},{\bf q},k) &\equiv & V_{11} ({\bf p},{\bf q},k)  + V_{12}({\bf p},{\bf q},k)+\ldots  + V_{1N} ({\bf p},{\bf q},k)\nonumber\\ 
&&{} + V_{21} ({\bf p},{\bf q},k)  + V_{22}({\bf p},{\bf q},k)+\ldots  + V_{2N} ({\bf p},{\bf q},k) \nonumber\\
&&{} + \cdots \nonumber\\
&&{} + V_{N1} ({\bf p},{\bf q},k)  + V_{N2}({\bf p},{\bf q},k)+\ldots  + V_{NN} ({\bf p},{\bf q},k)  \nonumber\\ 
&\equiv&  \left(
\begin{array}{cccc}
 1, & 1, & \ldots , &1 \\
\end{array}
\right)
\left(
\begin{array}{cccc}
 V_{11}({\bf p},{\bf q},k) , & V_{12}({\bf p},{\bf q},k), & \ldots & V_{1N}({\bf p},{\bf q},k) \\
 V_{21}({\bf p},{\bf q},k) , & V_{22}({\bf p},{\bf q},k), & \ldots & V_{2N}({\bf p},{\bf q},k) \\
 \ldots  & \ldots & \ldots & \ldots \\
 V_{N1}({\bf p},{\bf q},k) , & V_{N2}({\bf p},{\bf q},k), & \ldots & V_{NN}({\bf p},{\bf q},k) 
\end{array}
\right) \left(
\begin{array}{c}
1 \\
1 \\
\ldots \\
1
\end{array}
\right)\nonumber\\
&\equiv& \bar {\cal U} \,{\cal V}({\bf p},{\bf q},k)\, {\cal U}.
\label{eq2}
\end{eqnarray}
Similarly to the potential, we represent the scattering amplitude as 
\begin{eqnarray}
T({\bf p},{\bf q},k) &=& 
T_{11} ({\bf p},{\bf q},k)  + T_{12}({\bf p},{\bf q},k)+\ldots  + T_{1N} ({\bf p},{\bf q},k)\nonumber\\ 
&&{} + T_{21} ({\bf p},{\bf q},k)  + T_{22}({\bf p},{\bf q},k)+\ldots  + T_{2N} ({\bf p},{\bf q},k) \nonumber\\ 
&&{} +  \cdots  \nonumber\\ 
&&{} + T_{N1} ({\bf p},{\bf q},k)  + T_{N2}({\bf p},{\bf q},k)+\ldots  + T_{NN} ({\bf p},{\bf q},k)  \nonumber\\ 
&\equiv&  \left(
\begin{array}{cccc}
 1, & 1, & \ldots , &1 \\
\end{array}
\right)
\left(
\begin{array}{cccc}
 T_{11}({\bf p},{\bf q},k) , & T_{12}({\bf p},{\bf q},k), & \ldots & T_{1N}({\bf p},{\bf q},k) \\
 T_{21}({\bf p},{\bf q},k) , & T_{22}({\bf p},{\bf q},k), & \ldots & T_{2N}({\bf p},{\bf q},k) \\
 \ldots , & \ldots,& \ldots & \ldots \\
 T_{N1}({\bf p},{\bf q},k) , & T_{N2}({\bf p},{\bf q},k), & \ldots & T_{NN}({\bf p},{\bf q},k) 
\end{array}
\right) \left(
\begin{array}{c}
1 \\
1 \\
\ldots \\
1
\end{array}
\right)\nonumber\\
&\equiv& \bar {\cal U} \,{\cal T}({\bf p},{\bf q},k)\, {\cal U}.
\label{eq3}
\end{eqnarray}
We substitute Eqs.~(\ref{eq2}) and (\ref{eq3}) in Eq.~(\ref{eq1}) and,
removing $\bar {\cal U}$ and $ {\cal U}$ corresponding to the initial
and final states, demand that the following
matrix equation is satisfied 
\begin{equation}
{\cal T}({\bf p},{\bf q},k) = {\cal V}({\bf p},{\bf q},k)  +\int
\frac{d^3{\bf  l}}{(2\pi)^3} \, {\cal V}({\bf p},{\bf l},k) \, {\cal
  U} \, G(k, {\bf l}) \, \bar{\cal U} \, {\cal T}({\bf l},{\bf q},k).  
\label{eq4}
\end{equation}
Next, we introduce the cutoff-dependent potential via
\begin{eqnarray}
 V({\bf p},{\bf q},k,\bar \Lambda,\Lambda) &=& V_{11}({\bf p},{\bf q},k, \bar \Lambda,\Lambda) \,
                                  \theta(\bar\Lambda_1-p)
                                  \theta(\Lambda_1-q) + V_{12}({\bf
                                  p},{\bf q},k, \bar \Lambda,\Lambda) \,
                                  \theta(\bar\Lambda_1-p)
                                  \theta(\Lambda_2-q)  \nonumber\\ 
& + & \ldots + V_{1N}({\bf p},{\bf q},k, \bar \Lambda,\Lambda)\,\theta(\bar\Lambda_1-p) \theta(\Lambda_N-q)\nonumber\\
& + & V_{21}({\bf p},{\bf q},k, \bar \Lambda,\Lambda) \, \theta(\bar\Lambda_2-p)
      \theta(\Lambda_1-q) + V_{22}({\bf p},{\bf q},k, \bar \Lambda,\Lambda) \,
      \theta(\bar\Lambda_2-p) \theta(\Lambda_2-q)  \nonumber\\ 
& + & \ldots + V_{2N}({\bf p},{\bf q},k, \bar \Lambda,\Lambda)\,\theta(\bar\Lambda_2-p) \theta(\Lambda_N-q)\nonumber\\
& + & \cdots \nonumber\\
& + & V_{N1}({\bf p},{\bf q},k, \bar \Lambda,\Lambda) \, \theta(\bar\Lambda_{N}-p)
      \theta(\Lambda_1-q) + V_{N2}({\bf p},{\bf q},k, \bar \Lambda,\Lambda) \,
      \theta(\bar\Lambda_N-p) \theta(\Lambda_2-q)  \nonumber\\ 
& + & \ldots + V_{NN}({\bf p},{\bf q},k, \bar \Lambda,\Lambda)\,\theta(\bar\Lambda_N-p) \theta(\Lambda_N-q)\nonumber\\
&\equiv&   
 \left(
\begin{array}{cccc}
\theta( \bar\Lambda_1-p), & \theta(\bar\Lambda_2-p), & \ldots , &\theta( \bar\Lambda_N-p) \\
\end{array}
\right)\nonumber\\
&\times& 
\left(
\begin{array}{cccc}
 V_{11}({\bf p},{\bf q},k, \bar \Lambda,\Lambda) , & V_{12}({\bf
                                                     p},{\bf q},k,
                                                     \bar
                                                     \Lambda,\Lambda),
  & \ldots & V_{1N}({\bf p},{\bf q},k, \bar \Lambda,\Lambda) \\ 
 V_{21}({\bf p},{\bf q},k, \bar \Lambda,\Lambda) , & V_{22}({\bf
                                                     p},{\bf q},k,
                                                     \bar
                                                     \Lambda,\Lambda),
  & \ldots & V_{2N}({\bf p},{\bf q},k, \bar \Lambda,\Lambda) \\ 
 \ldots  & \ldots & \ldots & \ldots \\
 V_{N1}({\bf p},{\bf q},k, \bar \Lambda,\Lambda) , & V_{N2}({\bf
                                                     p},{\bf q},k,
                                                     \bar
                                                     \Lambda,\Lambda),
  & \ldots & V_{NN}({\bf p},{\bf q},k, \bar \Lambda,\Lambda)  
\end{array}
\right)
 \left(
\begin{array}{c}
\theta(\Lambda_1-q) \\
\theta(\Lambda_2-q) \\
\ldots \\
\theta(\Lambda_N-q)
\end{array}
\right)\nonumber\\
&\equiv & \bar\Theta(p){\cal V} ({\bf p},{\bf q},k, \bar \Lambda,\Lambda)\Theta(q) ,
\label{eq5}
\end{eqnarray}
where $\Lambda \equiv \{ \Lambda_i \}$, $\bar \Lambda \equiv \{ \bar
\Lambda_i \}$, $p \equiv |{\bf p}|$, $q \equiv |{\bf q}|$ and
$\theta(x)$ is the Heaviside theta function\footnote{While we use the
  sharp cutoff, our results are equally applicable for the theta
  functions replaced by smooth regulator functions.}, by
requiring that  
it satisfies the matrix equation 
\begin{equation}
 {\cal V} ({\bf p},{\bf q},k, \bar \Lambda, \Lambda)  = {\cal V} ({\bf
   p},{\bf q},k) + \int \frac{d^3{\bf  l}}{(2\pi)^3}\,  {\cal V} ({\bf
   p},{\bf l},k)  \left[  
{\cal U}  G(k, {\bf l}) \bar {\cal U}  -   
\Theta (l) G(k, {\bf l}) \bar\Theta (l)
\right] {\cal V} ({\bf l},{\bf q},k, \bar \Lambda , \Lambda). 
\label{eq6}
\end{equation}
It then follows from Eqs.~(\ref{eq4}) and (\ref{eq6}) that the
off-shell low-energy T-matrix ${\cal T}({\bf p},{\bf q},k)$ can be
obtained by solving the following equation 
\begin{equation}
{\cal T}({\bf p},{\bf q},k) = {\cal V}({\bf p},{\bf q},k, \bar \Lambda
,\Lambda)
+\int \frac{d^3{\bf  l}}{(2\pi)^3} \, {\cal V}({\bf p},{\bf
  l},k, \bar \Lambda ,\Lambda) \, \Theta (l) \, G(k,{\bf l}) \, \bar\Theta (l) \,
{\cal T}({\bf l},{\bf q},k).  
\label{eq7}
\end{equation}
Any solution of Eq.~(\ref{eq6}) also satisfies the following system of
$2 N$ RG equations
\begin{eqnarray}
\frac{\partial {\cal V} ({\bf p},{\bf q},k, \bar \Lambda ,\Lambda)}{\partial\Lambda_i}  &=& - \int \frac{d^3{\bf  l}}{(2\pi)^3}\,  {\cal V} ({\bf p},{\bf l},k, \bar \Lambda ,\Lambda)  \, \frac{\partial 
\left[ \Theta (l) G(k, {\bf l}) \bar\Theta (l)\right] }{\partial\Lambda_i} 
\,{\cal V} ({\bf l},{\bf q},k, \bar \Lambda ,\Lambda)\,,\nonumber \\ 
\frac{\partial {\cal V} ({\bf p},{\bf q},k, \bar \Lambda ,\Lambda)}{\partial \bar \Lambda_i}  &=& - \int \frac{d^3{\bf  l}}{(2\pi)^3}\,  {\cal V} ({\bf p},{\bf l},k, \bar \Lambda ,\Lambda)  \, \frac{\partial 
\left[ \Theta (l) G(k, {\bf l}) \bar\Theta (l)\right] }{\partial \bar \Lambda_i} 
\,{\cal V} ({\bf l},{\bf q},k, \bar \Lambda ,\Lambda)\,,
\label{RGeq}
\end{eqnarray}
with $i = 1, \ldots , N$. 
While the above derivation of Eq.~(\ref{RGeq}) served mainly for the
purpose of demonstrating of its physical content,  it can be directly
obtained from Eq.~(\ref{eq7}) by demanding cutoff independence of
${\cal T}({\bf p},{\bf q},k)$. Therefore, 
for ${\cal V}({\bf p},{\bf q},k, \bar \Lambda ,\Lambda)$, satisfying
Eqs.~(\ref{RGeq}), the off-shell amplitude  $T({\bf p},{\bf q},k)=\bar
{\cal U} \,{\cal T}({\bf p},{\bf q},k)\, {\cal U}$ obtained from the
solution of Eq.~(\ref{eq7}) is cutoff  
independent and coincides with the solution of Eq.~(\ref{eq1}) at low
energies, i.e. below all cutoffs $\Lambda_i$ and $\bar\Lambda_i$.  

The case of Hermitean cutoff-dependent potentials corresponds to
choosing $\bar \Lambda_i = \Lambda_i$ for all $i$, so that $\bar
\Theta (x) = \big(\Theta (x) \big)^{\rm T}$.   Furthermore, 
for a single cutoff parameter, Eq.~(\ref{RGeq}) reduces to the
differential RG equation of Ref.~\cite{Birse:1998dk}.  In general,
Eq.~(\ref{RGeq}) is a system of integro-differential equations,
however in some cases such as e.g.~for separable potentials, it can be
reduced to a system of differential equations.   

\section{RG equation with two cutoffs}
\label{RG2C}

In exact analogy to the previous section, one can obtain 
a system of RG equations for the LS equation in partial wave basis  
\begin{equation}
T(p,q,k)=V(p,q,k)+\int d l \, V(p,l,k) G(k, l)\,T(l,q,k),
\label{eq1t}
\end{equation}
where $G(k,l) = m\,l^2/(2\pi^2 (k^2-l^2+i\,\epsilon))$. 
The corresponding cutoff regularized potential, defined analogously to
Eq.~(\ref{eq5}), satisfies the following system of RG equations 
\begin{equation}
\frac{\partial {\cal V} ({p},{q},k,\Lambda)}{\partial\Lambda_i}  = -
\int d l\,  {\cal V} ({p},{l},k,\Lambda)  \, \frac{\partial  
\left[ \Theta (l) G(k, {l}) \bar\Theta (l)\right] }{\partial\Lambda_i} 
\,{\cal V} ({l},{q},k,\Lambda). 
\label{RGeqPW}
\end{equation}
Here and in what follows, we restrict ourselves to the case of
Hermitean potentials. 

As a simple application, we solve the RG equations with two
cutoff parameters, $\Lambda_1=\bar\Lambda_1$ and
$\Lambda_2=\bar\Lambda_2  <\Lambda_1$, as a perturbative power series 
expansion  
in the small parameters, $p$, $q$, $k$ and $\Lambda_2$. 
Specifically, we consider the cutoff regularized potential of the form
\begin{eqnarray}
 V({p},{q},k,\Lambda) &=& V_{11}(k,\Lambda) \, \theta(\Lambda_1-p)
                          \theta(\Lambda_1-q) +
                          V_{12}({p},{q},k,\Lambda) \,
                          \theta(\Lambda_1-p) \theta(\Lambda_2-q)
                          \nonumber\\ 
& + & V_{21}({p},{q},k,\Lambda) \, \theta(\Lambda_2-p)
      \theta(\Lambda_1-q) + V_{22}({p},{q},k,\Lambda) \,
      \theta(\Lambda_2-p) \theta(\Lambda_2-q)  \nonumber\\ 
&\equiv&   
 \left(
\begin{array}{cc}
\theta( \Lambda_1-p), & \theta(\Lambda_2-p) \\
\end{array}
\right) \left(
\begin{array}{cc}
 V_{11}(k,\Lambda) , & V_{12}({p},{q},k,\Lambda) \\
 V_{21}({p},{q},k,\Lambda)  , & V_{22}({p},{q},k,\Lambda)    
\end{array}
\right) \left(
\begin{array}{c}
\theta(\Lambda_1-q) \\
\theta(\Lambda_2-q) 
\end{array}
\right)\nonumber\\
&\equiv & \left( \Theta(p) \right)^{\rm T}{\cal V}_2 ({p},{q},k,\Lambda)\Theta(q) ,
\label{eqV2C}
\end{eqnarray}
where $V_{21}({p},{q},k,\Lambda) =V_{12}({q},{p},k,\Lambda) $.
We look for ${\cal V}_2$ as a solution to Eq.~(\ref{RGeqPW}) in the
form of a perturbative expansion in small parameters 
\begin{eqnarray}
{\cal V}_{2} ({p},{q},k,\Lambda) & = & {\cal V}_{\rm LO}
                                       ({p},{q},k,\Lambda) +{\cal
                                       V}_{\rm NLO} ({p},{q},k,\Lambda) +\cdots \nonumber \\ 
&=& \left(
\begin{array}{cc}
 V_{11, \, \rm LO}(k,\Lambda) , & 0 \\
 0  , & 0    
\end{array}
\right)+\left(
\begin{array}{cc}
 V_{11, \, \rm NLO}(k,\Lambda) , & q^2 V_{12, \, \rm NLO}(k,\Lambda)\\
 p^2 V_{12, \, \rm NLO}(k,\Lambda) , & 0    
\end{array}
\right)
+\cdots .
\label{V0}
\end{eqnarray}
We note that $V_{22}$ only appears at NNLO. 
As the potential is presented as a series in $p$ and $q$, the coefficients $V_{ij, \, \rm LO}$, $V_{ij, \, \rm NLO} $, etc. do not depend on these variables. Substituting Eq.~(\ref{V0})  into Eq.~(\ref{RGeqPW}) and solving order-by-order we obtain 
\begin{eqnarray}
{\cal V}_{\rm LO} ({p},{q},k,\Lambda) &=& \frac{4 \pi^2}{m} \left(
\begin{array}{cc}
 \frac{1}{- k \ln \frac{\Lambda _1-k}{\Lambda _1+k}-2\Lambda
   _1+2 C_1\left(k^2\right)}, & 0 \\
 0, & 0 \\
\end{array}
\right)\,,
\label{V01}
\nonumber\\
{\cal V}_{\rm NLO} ({p},{q},k,\Lambda) &=&
\left(
\begin{array}{cc}
 \frac{2 C_2\left(k^2\right) \left(-k^3 \ln \frac{\Lambda
   _2-k}{\Lambda _2+k}-2 \Lambda _2 k^2-\frac{2
   \Lambda _2^3}{3}-C_3\left(k^2\right)\right)}{\left(k \ln
   \frac{\Lambda _1-k}{\Lambda _1+k}+2 \Lambda
   _1-2 C_1\left(k^2\right)\right){}^2}, & \frac{q^2
   C_2\left(k^2\right)}{k \ln \frac{\Lambda
   _1-k}{\Lambda _1+k}+2 \Lambda _1-2
   C_1\left(k^2\right)} \\
 \frac{p^2 C_2\left(k^2\right)}{k \ln \frac{\Lambda_1-k}{\Lambda _1+k}+2 \Lambda _1-2
   C_1\left(k^2\right)}, & 0 \\
\end{array}
\right) \,,
\nonumber\\
& \cdots&  \, .
\label{V0NLO}
\end{eqnarray}
The functions $C_1$, $C_2$ and $C_3$ are analytic at $k^2=0$,
i.e.~they can be written  as Taylor series
\begin{eqnarray}
C_1(k^2) &=& c_{10} + c_{11} k^2 + c_{12}k^4 + \cdots,\nonumber\\
C_2(k^2) &=& c_{20} + c_{21} k^2 + c_{12}k^4 + \cdots,\nonumber\\
C_3(k^2) &=& 0 + c_{31} k^2 + c_{32}k^4 + \cdots ,
\label{CTS} 
\end{eqnarray}
where we have taken $C_3$ with a vanishing constant term in order that
the NLO potential is indeed suppressed by powers of small parameters.
All parameters of the potential cannot be fixed by demanding that the
empirical on-shell scattering amplitude is reproduced. Therefore, we
set all coefficients to zero except for $c_{10}$, $c_{20}$ and
$c_{31}$.  
We fix the remaining constants $c_{10}$  and $c_{31}$ by matching to
the low-energy scattering 
amplitude parameterized in the form of the effective range expansion  
\begin{equation}
T(k,k,k) \equiv T(k)=-\frac{4 \pi}{m} \,
\frac{1}{-\frac{1}{a}+\frac{r}{2} k^2+\cdots -i k},
\label{ERE}
\end{equation}
with $m$ the particle mass, $a$ the scattering length and $r$ the effective range.
We write Eq.~(\ref{ERE}) as a perturtbative expansion valid both for the case of
a natural and unnaturally large scattering length,
\begin{equation}
T(k) =T_{\rm LO}+T_{\rm NLO}+\cdots ,
\label{TEXP}
\end{equation}
where
\begin{eqnarray}
T_{\rm LO}(k) &=& - \frac{4 \pi}{m} \, \frac{1}{-\frac{1}{a} - i k} \,
                  ,\nonumber\\
T_{\rm NLO}(k) &=&  \frac{2 \pi}{m} \, \frac{r \,k^2}{\left( -\frac{1}{a} - i k\right)^2} \,,\nonumber\\
 & \cdots & \, .
\label{CTS1} 
\end{eqnarray}
By demanding that $V_{\rm LO}$ reproduces $T_{\rm LO}$ and the
perturbative inclusion of $V_{\rm NLO}$ generates $T_{\rm NLO}$, we obtain
\begin{eqnarray}
c_{10} &=& \frac{\pi}{2\,a},\nonumber\\
c_{31} &=& -\frac{\pi }{a}-\frac{\pi ^3 r}{c_{20} m}.
\label{cs}
\end{eqnarray}
The coefficient $c_{20}$ remains undetermined and parameterizes the
remaining freedom in the choice of the off-shell potential. 

Substituting the obtained values of $c_{ij}$ back into the potential, we find
\begin{eqnarray}
{\cal V}_{\rm LO} ({p},{q},k,\Lambda) & = & \left(
\begin{array}{cc}
 \frac{4 a \pi ^2}{m \left(-a k \ln \frac{\Lambda
   _1-k}{\Lambda _1+k}-2 a \Lambda _1+\pi \right)}, & 0
   \\
 0, & 0 \\
\end{array}
\right)\,,
\nonumber\\
{\cal V}_{\rm NLO} ({p},{q},k,\Lambda) &=&
\left(
\begin{array}{cc}
 \frac{2 a \left[3 a k^2 \pi
   ^3 r-m c_{20} \left(2 a \Lambda _2^3+6 a k^2
   \Lambda _2+3 k^2 \left(a k \ln \frac{\Lambda
   _2-k}{\Lambda _2+k}-\pi \right)\right)\right]}{3 m \left(-a k \ln \frac{\Lambda
   _1-k}{k+\Lambda _1}-2 a \Lambda _1+\pi \right){}^2},
   & \frac{a q^2 c_{20}}{a k \ln \frac{\Lambda
   _1-k}{\Lambda _1+k}+2 a \Lambda _1-\pi } \\
 \frac{a p^2 c_{20}}{a k \ln \frac{\Lambda
   _1-k}{\Lambda _1+k}+2 a \Lambda _1-\pi }, & 0 \\
\end{array}
\right)\,, \nonumber\\
& \cdots& \, .
\label{V0NLOexpl}
\end{eqnarray}
For $\Lambda_1\sim \Lambda_{\rm hard}$ and $c_{20}\sim 1/\Lambda_{\rm
  hard}^3$, where $\Lambda_{\rm hard}$ denotes the hard scale of the
problem (i.e.~the pion mass for the case at hand), the potential
corresponding to Eq.~(\ref{V0NLOexpl}) satisfies Weinberg's power  
counting \cite{Weinberg:1991um} for a system being close to the
unitary limit, i.e.~for $a \gg  \Lambda_{\rm hard}^{-1}$. 
This differs from the results with a single cutoff parameter for which
the obtained power counting is the one of Ref.~\cite{Birse:1998dk}.

\section{Summary and conclusions}
\label{sec:summ}

Our paper provides a generalization of the Wilsonian renormalization
group approach to the Lippmann-Schwinger equation for two-particle
scattering at low energy by introducing a multitude of cutoff parameters. 
We derive a system of integro-diffrential equations for the cutoff regularized
potential, which reduces to the RG equation of
Ref.~\cite{Birse:1998dk} for the case of a single cutoff. 
As a simple application, we considered a perturbative solution of the
system of RG equations in the form of a power series  
expansion in momenta and energy. We have demonstrated that by introducing two cutoff parameters, 
one obtains a perturbative expansion of the potential which follows
the Weinberg power counting rules \cite{Weinberg:1991um}, while, as
shown in Ref.~\cite{Birse:1998dk}, the usage of a single cutoff parameter leads to
the power counting of Refs.~\cite{Kaplan:1998tg,vanKolck:1997ut,vanKolck:1998bw}.
This simple example demonstrates that the enlargement of the space of
the renormalization group parameters by exploiting the full freedom in
the choice of renormalization conditions can be advantageously used in the
context of the low-energy EFT for nucleon-nucleon scattering. 
It will be intersting to apply the presented formalism with the multitude of cutoff
parameters to the case of the potentials with a long-range
interaction. This work is in progress.

\acknowledgments
EE and JG  are grateful to M.~Birse for useful discussions on the Wilsonian RG
approach. 
This work was supported in part by BMBF (contract No.~05P2015 - NUSTAR
R\&D), by DFG and NSFC through funds provided to the
Sino-German CRC 110 ``Symmetries and the Emergence of Structure in QCD" (NSFC
Grant No.~11621131001, DFG Grant No.~TRR110), by the Georgian Shota Rustaveli National
Science Foundation (grant FR/417/6-100/14) and by the CAS President's International
Fellowship Initiative (PIFI) (Grant No.~2017VMA0025).


\end{document}